\documentclass[12pt]{article}
\usepackage{epsfig}
\usepackage{subfigure}
\usepackage{amsmath}
\usepackage{amsfonts}
\usepackage{amssymb}
\usepackage{url}
\usepackage{hyperref}
\newcommand{\be}{\begin{equation}}
\newcommand{\ee}{\end{equation}}
\newcommand{\bea}{\begin{eqnarray}}
\newcommand{\eea}{\end{eqnarray}}

\begin{document}
\begin{titlepage}
\begin{flushright}
RITS-PP-004\\
KEK-TH-1082\\
\end{flushright}
\vspace{4\baselineskip}
\begin{center}
{\Large\bf D-term Contributions to the Mixed Modulus-Anomaly 
 Mediated Supersymmetry Breaking}
\end{center}
\vspace{1cm}
\begin{center}
{\large Takeshi Fukuyama$^{a,}$
\footnote{\tt E-mail:fukuyama@se.ritsumei.ac.jp},
Tatsuru Kikuchi$^{a,b}$
\footnote{\tt E-mail:tatsuru@post.kek.jp}
and Nobuchika Okada$^{b,c,}$
\footnote{\tt E-mail:okadan@post.kek.jp}}
\end{center}
\vspace{0.2cm}
\begin{center}
${}^{a}$ {\small \it Department of Physics, Ritsumeikan University,
Kusatsu, Shiga, 525-8577, Japan}\\[.2cm]
${}^{b}$ {\small \it Theory Division, KEK,
Oho 1-1, Tsukuba, Ibaraki, 305-0801, Japan}\\
${}^{c}$ {\small \it Department of Particle and Nuclear Physics,
The Graduate University for Advanced Studies, \\
Oho 1-1, Tsukuba, Ibaraki, 305-0801, Japan}\\
\medskip
\vskip 10mm
\end{center}
\vskip 10mm
\begin{abstract}
We investigate effects of $D$-term contributions 
 to the mixed modulus-anomaly mediated supersymmetry 
 breaking scenario. 
In the original scenario, the tachyonic slepton problem 
 in the pure anomaly mediated supersymmetry breaking 
 is cured by modulus contributions. 
We generalize the scenario 
 so as to include contributions from the $D$-terms 
 of $U(1)_Y$ and the gauged $U(1)_{\rm B-L}$ 
 which is motivated in a grand unified theory 
 based on a higher rank gauge group such as SO(10). 
As a consequence of additional $D$-term contributions 
 to scalar masses, we obtain various soft supersymmetry 
 breaking mass spectra, which are different from 
 those obtained in the conventional mixed modulus-anomaly 
 mediated supersymmetry breaking. 
Especially, we find that the lightest superpartner (LSP) neutralinos 
 can be various types, such as Higgsino-like, wino-like and bino-like 
 degenerating with the next LSP sfermions. 
\end{abstract}
\end{titlepage}
\section{Introduction}

Supersymmetry (SUSY) extension is one of the most promising way 
 to solve the gauge hierarchy problem in the standard model \cite{SUSY}. 
However, since any superpartners have not been observed 
 in current experiments, SUSY should be broken at low energies. 
Furthermore, soft SUSY breaking terms are severely constrained 
 to be almost flavor blind and CP invariant. 
Thus, the SUSY breaking has to be mediated to the visible sector 
 in some clever way not to induce too large CP and flavor violation effects. 
Some mechanisms to achieve such SUSY breaking mediations  
 have been proposed \cite{Luty:2005sn}. 

The anomaly mediated supersymmetry breaking (AMSB) 
 \cite{AMSB1, AMSB2} is one of the most attractive scenario 
 due to its flavor-blindness and ultraviolet (UV) insensitivity 
 for the resultant soft SUSY breaking terms. 
Unfortunately, the pure AMSB scenario is obviously excluded, 
 since it predicts slepton squared masses being negative. 
There have been many attempts to solve this problem 
 by taking into account additional positive contributions 
 to slepton squared masses 
 at tree level \cite{AMSB1, tree1, tree2} 
 or at quantum level \cite{PR, quantum}. 
Among them, 
 adding $D$-terms of the $U(1)_Y$ and the (gauged) $U(1)_{\rm B-L}$ 
 may be the most interesting possibility, 
 because these $U(1)$ symmetries are anomaly-free 
 with respect to the standard model gauge group 
 so that the UV insensitivity is preserved \cite{tree1}. 
We can expect such new contributions from the $D$-terms, 
 if some grand unified theory (GUT) based on 
 a higher rank gauge group such as SO(10) 
 takes place at high energies, 
 which includes new Higgs fields and 
 the gauged $U(1)_{\rm B-L}$ as its subgroup. 
However, 
 it has been found that this scenario requires 
 a very small $\tan\beta$ \cite{Kitano:2004zd} 
 to obtain the correct electroweak symmetry breaking. 
As a result, the top Yukawa coupling blows up 
 far below the GUT scale, 
 and the minimal supersymmetric standard model (MSSM) 
 cannot be simply connecting into SUSY GUTs 
 in the way usually expected. 

Recently, Kachru-Kallosh-Linde-Trivedi (KKLT) \cite{Kachru:2003aw} 
 have proposed a way to stabilize the modulus 
 in string theories with flux compactification. 
Interestingly, a stabilized modulus can induce 
 additional SUSY breaking contributions 
 comparable to the pure AMSB contributions, 
 so as to solve tachyonic slepton problem. 
There have already been several studies 
 on the SUSY breaking mediation 
 in this KKLT type setup \cite{Choi:2004sx, KKLT2, KKLT3}, 
 the so-called mixed modulus-anomaly mediation, 
 and the characteristic sparticle mass spectrum  
 have been obtained. 

In this paper, we generalize 
 the mixed modulus-anomaly mediation scenario 
 so as to include the effects of the $D$-term contributions. 
Contributions from the mixed modulus-anomaly mediation 
 play a role to widen the allowed region of $\tan \beta$, 
 so that the top Yukawa coupling remains perturbative 
 until the GUT scale and the MSSM can simply connect into GUTs. 
On the other hand, 
 the $D$-term contributions change sfermion mass spectrum 
 from the one in the conventional mixed modulus-anomaly 
 mediation scenario. 
As a result, the sparticle mass spectrum in our scenario 
 can be quite different from the one obtained 
 in the mixed modulus-anomaly mediation scenario 
 or in the AMSB scenario with the $D$-term contributions. 

This paper is organized as follows. 
In section 2, we briefly review on the KKLT setup. 
In section 3, we give formulas of the mixed modulus-anomaly mediation  
 including $D$-term contributions from $U(1)_Y$ and $U(1)_{\rm B-L}$, 
 which are necessary for our numerical analysis. 
The results of numerical analysis are presented in section 4. 
The last section is devoted to summary and discussions.

\section{KKLT setup}

In this section, we work out in the superconformal framework 
 of supergravity \cite{Kaku:1978nz}. 
A modulus superfield $T$ plays a crucial role in the KKLT setup, 
 whose basic Lagrangian is given by 
\bea
 {\cal L} &=& - 3 \int d^4 \theta \;\phi^\dag \phi \; e^{-K/3}  
  + \int d^2 \theta \;\phi^3 W + h.c. \; , 
\eea
where $\phi=1+ \theta^2 F_{\phi}$ is the compensating multiplet. 
Here, the K\"ahler potential is taken to be the no-scale type, 
\bea
K = -3 \ln (T + T^\dagger) \; ,  
\eea
and the following superpotential is derived 
 in the context of the type IIB string theory \cite{Kachru:2003aw}, 
\bea
W = W_0 -C e^{-a T} \; ,
\label{KKLT}
\eea
where the first term is a constant, and the second term is generated 
 through the $SU(N_c)$ gaugino condensation 
 with coefficients $C$ and $a=8 \pi^2/N_c$ 
 being real and positive 
\footnote{
 In the original work by KKLT \cite{Kachru:2003aw}, 
 $C=1$ and $a=0.1$ were used 
 in order to realize de Sitter (or Minkowski) vacua.}. 

With these K\"ahler potential and superpotential, 
 the scalar potential is given by 
\bea
V= \frac{T+T^\dagger}{3} |W_T|^2 - W W_T^\dagger -W^\dagger W_T 
 \;  ,
\eea 
where $W_T = \partial W/\partial T$. 
This scalar potential has a supersymmetric anti-de Sitter minimum,  
\bea
V = - 3 |F_{\phi}|^2 \left( T + T^\dagger + \frac{2}{a} \right)  < 0 
 \; , 
\eea
 with $F_\phi = W_T^\dagger/3$. 
At the potential minimum, the $F$-term of the modulus 
 is given by 
\be
 F_T = \frac{2}{a} F_\phi =\frac{N_C}{4 \pi^2} F_\phi \; . 
\label{FT}
\ee
In order to obtain a de Sitter (or Minkowski) vacuum, 
 the lifting potential due to the presence of the anti-D3 brane 
 is introduced \cite{Kachru:2003aw},  
\bea
\Delta V = \frac{D}{(T + T^\dagger)^n} \; ,
\eea
where $n$ is an integer ($n=2$ in the original KKLT paper),  
 and $D$ is a constant whose value is tuned 
 so as to realize the de Sitter (or Minkowski) vacuum. 
At the de Sitter (or Minkowski) vacuum, 
 only $\Re [T]$ of $T$ has non-zero vacuum expectation value  
 and the relation $F_T \simeq F_\phi/a =N_c F_T/(4 \pi^2)$ 
 in Eq.~(\ref{FT}) still holds. 

Next, let us introduce the MSSM sector into the modulus Lagrangian. 
The K\"ahler potential is replaced to the one 
 including the MSSM matter and Higgs superfields, 
 $K(T, T^\dagger) \rightarrow K(T,T^\dagger) + K_{MSSM}$. 
For simplicity, we take the minimal K\"ahler potential 
 for the MSSM superfields, $K_{MSSM} = Q_i^\dagger e^{2g_a V_a} Q_i$, 
 where $Q_i$ stands for the MSSM matter and Higgs superfields. 
Expanding $e^{K/3}$, 
 the K\"ahler potential for the MSSM superfields is described as 
\bea 
 \int d^4 \theta \phi^\dagger \phi 
 \left( T+T^\dagger \right)  
 Q_i^\dagger e^{2g_a V_a} Q_i  + \cdots \; .
\label{matterK}
\eea
For the gauge sector in the MSSM, the kinetic term is of the form, 
\bea
{\cal L}_{gauge} = \frac{1}{4} \int d^2 \theta 
 f_a  {\cal W}^{a \alpha} {\cal W}^a_\alpha \; . 
\label{gaugeK}
\eea
We take the gauge kinetic function $f_a=T$ in the following. 

In the above setup, there are two SUSY breaking sources, 
 namely $F_\phi$ and $F_T$. 
Non-zero $F_\phi$ induces soft SUSY breaking terms 
 through the AMSB, and the resultant SUSY breaking mass scale is 
 characterized by $m_{AMSB} \sim F_\phi/(16 \pi^2)$. 
On the other hand, as can be easily seen 
 from Eqs.~(\ref{matterK}) and (\ref{gaugeK}), 
 non-zero $F_T$ leads to soft SUSY breaking terms at tree level, 
 the modulus mediation. 
The resultant SUSY breaking mass scale in the modulus mediation 
 is characterized by 
\bea 
 m_{modulus} \sim \frac{F_T}{T+T^\dagger} \; . 
\eea
Noting $\Re[T]=1/g_{\rm GUT}^2 ={\cal O}(1)$ 
 ($g_{\rm GUT}$ denotes the standard model gauge coupling 
 at the GUT scale) 
 and $F_T \simeq N_C F_\phi/(4 \pi^2)$, 
 we see that this contribution by the moduli mediation 
 is comparable to the one by the AMSB, $m_{AMSB} \sim m_{modulus}$. 
This fact is the key of the mixed modulus-anomaly mediation scenario.
According to the method developed in Ref.~\cite{method} 
 (see also Ref.~\cite{PR}), 
 soft SUSY breaking terms 
 (each gaugino masses $M_a$, sfermion squared masses $\widetilde{m}_i^2$ 
 and $A$-parameters) 
 at the GUT scale ($\mu=M_{\rm GUT}\simeq 2 \times 10^{16}$ GeV)\footnote{
In this paper, we set the compactification scale of 
 the string theory to be the GUT scale.}
 can be extracted from 
 renormalized gauge kinetic functions and 
 SUSY wave function renormalization coefficients \cite{Choi:2004sx}, 
\bea 
&& M_{a} = M (\alpha + b_a g_a^2) \; , 
 \nonumber\\ 
&& \tilde{m}_i^2 = M^2 
     \left( 
    \alpha^2 + 2 \alpha (T + T^\dagger) \partial_{T} \gamma_i  
     - 8 \pi^2 \mu \partial_\mu  \gamma_i  \right) \; , 
 \nonumber \\
&& A_{ijk} = M \left( 
  3 \alpha - \gamma_i -\gamma_j - \gamma_k \right)\; .
\label{softterms}
\eea
Here, $g_a$ ($g_1=g_2=g_3=g_{\rm GUT}$) are the gauge couplings, 
      $b_a$ are the beta function coefficients, and 
   $\gamma_i$ are the anomalous dimensions 
   which depend on $T$ through the T-dependence of 
   the gauge couplings and Yukawa couplings. 
Parameters $M$ (typical soft SUSY breaking mass scale) and 
 $\alpha$ are defined as 
\bea 
&&M = \frac{F_\phi}{16 \pi^2} \sim \frac{m_{3/2}}{16 \pi^2} \;, \\
&&\alpha M = \frac{F_T}{T+T^\dagger}
\eea  
with the gravitino mass $m_{3/2}$. 
The results of the pure AMSB is reproduced 
 in the limit $\alpha \rightarrow 0$, 
 while the limit $\alpha \gg 1 $ corresponds to 
 the pure modulus mediation 
 whose contribution to sfermion masses is positive. 
As discussed above, $\alpha = {\cal O}(1)$ is expected 
 in the KKLT setup, 
 so that both the AMSB and the moduli mediation 
 give important contributions 
 to resultant soft SUSY breaking parameters. 

There are remaining two parameters in the Higgs sector, 
 namely $\mu$ and $B \mu$ terms, 
 that are responsible for electroweak symmetry breaking
 and should be of the order of the electroweak scale. 
As in the AMSB scenario, 
 the natural value of the $B$-parameter would be 
 $B \sim m_{3/2} \gg M$, and 
 the Higgs sector should be extended in order to 
 achieve the $B$-parameter being at the electroweak scale. 
Although some fine-tuning among parameters is necessary, 
 the way to realize $\mu \sim B \sim M$ 
 have been discussed in Ref.~\cite{Choi:2004sx}. 
In our analysis, we treat them as free parameters as usual, 
 that is, $\mu$ and $B \mu$ are replaced 
 into two free parameters $\tan \beta$ and ${\rm sgn}(\mu)$, 
 while the value of $|\mu|$ is determined by the stationary condition 
 of the Higgs potential. 
In the next section, we consider to add two $D$-terms 
 of $U(1)_{Y}$ and $U(1)_{\rm B-L}$, and hence 
 total set of free parameters in our analysis is 
\be
\{ M,~\alpha,~D_Y,~D_{\rm B-L},~\tan \beta,~{\rm sgn}(\mu) \}\;.
\ee
%

\section{Mixed modulus-anomaly mediation including D-terms}

Now let us introduce the $D$-terms 
 to the mixed modulus-anomaly mediation. 
If there exists a $U(1)$ gauge multiplet 
 having a non-zero $D$-term, 
 the kinetic term of a matter superfield gives 
\be
{\cal L} = \int d^4 \theta \, Q_i^\dag e^{q_i V} Q_i
 \supset q_i D\, \widetilde{Q}_i^\dagger 
 \widetilde{Q}_i \;,
\ee
where $q_i$ is the $U(1)$ charge of the chiral multiplet $Q_i$. 
This leads to a shift for the scalar squared mass, 
\bea
\widetilde{m}_i^2 \to \widetilde{m}_i^2 - q_i D \;.
\eea
The $U(1)$ symmetry providing the $D$-term should be anomaly-free 
 in order not to induce quadratic divergence in a theory. 
As such a $U(1)$ symmetry, there exist two candidates in the MSSM, 
 namely $U(1)_{Y}$ and gauged $U(1)_{\rm B-L}$. 
Introduction of this $U(1)_{\rm B-L}$ gauge symmetry 
 is well-motivated, 
 if we assume that the MSSM is embedded into 
 a GUT based on a higher rank gauge group such as SO(10) 
 which includes the gauged $U(1)_{\rm B-L}$ as a subgroup. 
This possibility is our motivation to consider the $D$-term 
 in addition to the mixed modulus-anomaly mediation. 
Normally, many extra Higgs fields are involved in such models, 
 and some of them have non-zero vacuum expectation values 
 to break the GUT symmetry at the supersymmetric level. 
Once soft SUSY breaking terms for these Higgs fields 
 are taken into account, 
 the vacuum would be realized at the point 
 slightly away from the D-flat directions, 
 so that non-zero $D$-terms are developed. 
Although it depends on the detailed structure 
 of the Higgs sector, 
 we may naturally expect the scale of the $D$-term 
 to be $D \sim M^2$. 

Calculating the anomalous dimensions \cite{Choi:2004sx}, 
 all the soft SUSY breaking terms can be obtained 
 from Eq.~(\ref{softterms}). 
Taking $U(1)_Y$ and $U(1)_{\rm B-L}$ $D$-term contributions into account, 
 the soft scalar masses for the first two generations 
 at the GUT scale are explicitly written as 
\bea
m_{\widetilde{q}_{1,2}}^2 &=&M^2
\left[\frac{157}{25} g_{\rm GUT}^4 
- \frac{42}{5} g_{\rm GUT}^2 \,\alpha + \alpha^2 
-\frac{1}{6} \alpha_{Y} - \frac{1}{3} \alpha_{\rm B-L} 
\right] \;,
\nonumber\\
m_{\widetilde{u^c}_{1,2}}^2 &=&M^2
\left[\frac{112}{25} g_{\rm GUT}^4 
- \frac{32}{5} g_{\rm GUT}^2 \,\alpha + \alpha^2 
+ \frac{2}{3} \alpha_{Y} + \frac{1}{3} \alpha_{\rm B-L} 
\right] \;,
\nonumber\\
m_{\widetilde{d^c}_{1,2}}^2 &=& M^2
\left[\frac{178}{25} g_{\rm GUT}^4 
- \frac{28}{5} g_{\rm GUT}^2 \,\alpha + \alpha^2 
- \frac{1}{3} \alpha_{Y} + \frac{1}{3} \alpha_{\rm B-L} 
\right] \;,
\nonumber\\
m_{\widetilde{\ell}_{1,2}}^2 &=&M^2 
\left[-\frac{87}{25} g_{\rm GUT}^4 
- \frac{18}{5} g_{\rm GUT}^2 \,\alpha + \alpha^2 
+ \frac{1}{2} \alpha_{Y} + \alpha_{\rm B-L} \right] \;,
\nonumber\\
m_{\widetilde{e^c}_{1,2}}^2 &=& M^2 
\left[- \frac{198}{25} g_{\rm GUT}^4
- \frac{12}{5} g_{\rm GUT}^2 \,\alpha + \alpha^2 
- \alpha_{Y} - \alpha_{\rm B-L} \right]\;.
\eea
Here, we have defined $\alpha_Y$ and $\alpha_{\rm B-L}$ as 
\bea
\alpha_Y \equiv \frac{D_Y}{M^2} \; , ~~ 
\alpha_{\rm B-L} \equiv \frac{D_{\rm B-L}}{M^2} \; ,
\eea
and Yukawa couplings of the first two generations 
 have been neglected as a good approximation. 
For the third generation sfermion masses, 
 Yukawa couplings are involved, 
\bea
m_{\widetilde{q}_{3}}^2 &=& M^2
\left[\frac{157}{25} g_{\rm GUT}^4 + y_t^2 b_{y_t} + y_b^2 b_{y_b} 
- \left\{\frac{42}{5} g_{\rm GUT}^2 
- 6 \left(y_t^2 + y_b^2 \right) \right\} \alpha + \alpha^2 
- \frac{1}{6} \alpha_{Y} - \frac{1}{3} \alpha_{\rm B-L} \right] \;,
\nonumber\\
m_{\widetilde{u^c_3}}^2 &=& M^2
\left[\frac{112}{25} g_{\rm GUT}^4 + 2 y_t^2 b_{y_t} 
- \left(\frac{32}{5} g_{\rm GUT}^2 - 12 y_t^2 \right) \alpha + \alpha^2 
+ \frac{2}{3} \alpha_{Y} + \frac{1}{3} \alpha_{\rm B-L} \right] \;,
\nonumber\\
m_{\widetilde{d^c_3}}^2 &=& M^2
\left[\frac{178}{25} g_{\rm GUT}^4 + 2 y_b^2 b_{y_b}
- \left(\frac{28}{5} g_{\rm GUT}^2 - 12 y_b^2 \right) \alpha + \alpha^2
- \frac{1}{3} \alpha_{Y} + \frac{1}{3} \alpha_{\rm B-L} \right]\;,
\nonumber\\
m_{\widetilde{\ell}_{3}}^2 &=& M^2
\left[-\frac{87}{25} g_{\rm GUT}^4 + y_\tau^2 b_{y_\tau} 
- \left(\frac{18}{5} g_{\rm GUT}^2 - 6 y_\tau^2 \right) \alpha + \alpha^2 
+ \frac{1}{2} \alpha_{Y} + \alpha_{\rm B-L} \right] \;,
\nonumber\\
m_{\widetilde{e^c_3}}^2 &=&M^2
\left[- \frac{198}{25} g_{\rm GUT}^4 + 2 y_\tau^2 b_{y_\tau} 
- \left(\frac{12}{5} g_{\rm GUT}^2 - 12 y_\tau^2 \right) \alpha + \alpha^2 
- \alpha_{Y} - \alpha_{\rm B-L} \right] \;,
\eea
where $b_{y_t}$, $b_{y_b}$ and $b_{y_\tau}$ are given by 
\bea
b_{y_t} &=& 6 y_t^2 + y_b^2 - \frac{46}{5} g_{\rm GUT}^2 \;,
\nonumber\\
b_{y_b} &=& y_t^2 + 6 y_b^2 + y_\tau^2 - \frac{44}{5} g_{\rm GUT}^2 \;,
\nonumber\\
b_{y_\tau} &=& 3 y_b^2 + 4 y_\tau^2 - \frac{24}{5} g_{\rm GUT}^2 \;.
\eea
When the condition, $\alpha_Y < -\alpha_{\rm B-L} < 1/2 \alpha_Y$, 
 is satisfied with $\alpha_Y <0$ and $\alpha_{\rm B-L}>0$, 
 slepton squared masses obtain positive contributions 
 from $D$-terms. 
On the other hand, the $D$-terms in this region 
 give negative contributions to 
 $m_{\widetilde{q_i}}^2$ and $m_{\widetilde{u^c_i}}^2$, 
 while positive to $m_{\widetilde{d^c_i}}^2$. 

$A$-parameters are given by 
\bea
 A_{ijk} = M \left( 
  3 \alpha - \gamma_i -\gamma_j - \gamma_k \right)\; 
\eea 
 with explicit formulas of the anomalous dimensions, 
\bea
 \gamma_{q_i} &=& \frac{21}{5} g_{\rm GUT}^2-( y_t^2+y_b^2) 
 \delta_{i 3} \;, \nonumber \\
 \gamma_{u^c_i}&=&\frac{16}{5} g_{\rm GUT}^2- 2 y_t^2 
 \delta_{i 3} \; ,  \nonumber \\
 \gamma_{d^c_i} &=& \frac{14}{5} g_{\rm GUT}^2- 2 y_b^2 
 \delta_{i 3} \; ,  \nonumber \\
 \gamma_{\ell_i} &=& \frac{9}{5} g_{\rm GUT}^2-  y_\tau^2 
 \delta_{i 3} \; ,  \nonumber \\
 \gamma_{e^c_i} &=& \frac{6}{5} g_{\rm GUT}^2- 2 y_\tau^2 
 \delta_{i 3} \; ,  \nonumber \\
 \gamma_{H_1} &=& \frac{9}{5} g_{\rm GUT}^2- 3 y_b^2 -y_\tau^2 
 \; , \nonumber \\
 \gamma_{H_2} &=& \frac{9}{5} g_{\rm GUT}^2- 3 y_t^2 
 \; .
\eea

Also, the Higgs soft masses at the GUT scale are given by
\bea
m_{H_1}^2 &=& M^2
\left[-\frac{87}{25} g_{\rm GUT}^4 + 3 y_b^2 b_{y_b} + y_\tau^2 b_{y_\tau} 
- \left(\frac{18}{5} g_{\rm GUT}^2 - 18 y_b^2 - 6 y_\tau^2 \right) \alpha 
+ \alpha^2 + \frac{1}{2} \alpha_{Y} \right] \;,
\nonumber\\
m_{H_2}^2 &=& M^2
\left[-\frac{87}{25} g_{\rm GUT}^4 + 3 y_t^2 b_{y_t} 
- \left(\frac{18}{5} g_{\rm GUT}^2 - 18 y_t^2 \right) \alpha 
+ \alpha^2 - \frac{1}{2} \alpha_{Y} \right] \;.
\eea
The Higgs mass parameters, $\mu$-term and $B \mu$-term, 
 are determined from the electroweak symmetry breaking conditions, 
\bea
\mu^2 &=& 
\frac{m_{H_1}^2 - m_{H_2}^2 \tan^2 \beta}{\tan^2 \beta -1}
- \frac{1}{2} M_Z^2  \;,
 \nonumber\\
B \mu &=& \frac{1}{2} 
 \left[m_{H_1}^2 + m_{H_2}^2  + 2 \mu^2 \right] \sin 2 \beta \; .
\eea
In numerical analysis presented in the next section, 
 we found that the condition to provide the correct 
 electroweak symmetry breaking gives the most severe constraint 
 on the parameter space $(\alpha,~\alpha_{Y},~\alpha_{\rm B-L})$, 
 rather than that to provide non-tachyonic sfermion masses 
 $m_{\tilde{f}}^2 >0$. 

Finally, we give the explicit formulas of 
 the gaugino masses at the GUT scale, 
\bea
M_1 &=& M \left(\alpha + \frac{33}{5} g_{\rm GUT}^2 \right) \; , 
 \nonumber\\
M_2 &=& M \left(\alpha + g_{\rm GUT}^2 \right) \; ,
 \nonumber\\
M_3 &=& M \left(\alpha - 3 g_{\rm GUT}^2 \right) \; .
\eea

Inputting the soft SUSY breaking terms expressed above 
 at the GUT scale as the boundary conditions, 
 the soft SUSY breaking terms at the electroweak scale 
 are obtained through the renormalization group equations (RGEs). 
In the next section, we show 
 the resultant soft SUSY breaking mass spectrum 
 for various inputs of $M$, $\alpha$, $\alpha_Y$ 
 and $\alpha_{\rm B-L}$ with a given $\tan \beta$.

\section{Numerical results}
Now we are ready to perform a numerical evaluation 
 by using the formulas presented in the previous sections. 
With given $\tan \beta$ and the parameter set 
 ($\alpha$, $\alpha_Y$, $\alpha_{\rm B-L}$ ), 
 we input the formulas for soft SUSY breaking terms at the GUT scale, 
 and then evolve them according to the one-loop RGEs \cite{RGE}. 
In our analysis, we take an averaged soft SUSY breaking mass scale 
 as $M_S = 500$ GeV and evaluate all the soft SUSY breaking parameters 
 at this scale. 
As examples, we investigate the cases of $\tan \beta = 10$ 
 and $\tan \beta =45$ 
 with the unified gauge coupling constant 
 $\alpha_{\rm GUT}^{-1} = 25.4$ at $M_{\rm GUT}$. 
As a good approximation, we consider Yukawa couplings 
 only for fermions in the third generation 
 with input values at $M_{\rm GUT}$ as 
\bea
y_t = 0.635 \;,~~y_b = 0.0616\;,~~y_\tau = 0.0687\;,
\eea
for $\tan \beta = 10$ 
and 
\bea
y_t = 0.749 \;,~~y_b = 0.449 \;,~~y_\tau = 0.454 \;,
\eea 
for $\tan \beta = 45$, respectively.

First we examine the allowed region of the parameter space 
 ($\alpha$, $\alpha_Y$, $\alpha_{\rm B-L}$) 
 for given $\tan \beta=10$ and $45$ and $M=500$ GeV. 
Sparticle mass spectrum for various inputs in the range of 
 $0 \leq \alpha \leq 6$ and 
 $-10 \leq \alpha_Y, \alpha_{\rm B-L} \leq 10$ 
 has been calculated in every $0.2$ intervals for $\alpha$ 
 and in every $0.5$ intervals for $\alpha_Y$ and $\alpha_{\rm B-L}$. 
The allowed parameter sets of 
 ($\alpha, \alpha_Y$) and ($\alpha, \alpha_{\rm B-L}$) 
 are plotted in Fig.~1 and 2, 
 for which resultant sfermion squared masses are all positive 
 and the electroweak symmetry breaking is correctly achieved. 
In both $\tan \beta=10$ and $45$ cases, 
 the allowed region is severely constrained 
 for $\alpha \lesssim 2$ mainly due to the condition 
 for the correct electroweak symmetry breaking. 
In particular, for a large $\tan \beta$, 
 the soft mass squared of the down-type Higgs doublet 
 is likely to be $m_{H_1} \lesssim m_{H_2} < 0$, 
 so that it becomes difficult to achieve 
 the correct electroweak symmetry breaking.

We have performed the same analysis in the case of 
 the conventional mixed modulus-anomaly mediation 
 ($\alpha_Y = \alpha_{\rm B-L} =0$), 
 and found that the allowed region is constrained 
 to be $\alpha \gtrsim 2.5$ for $\tan \beta =10$ 
 and $\alpha \gtrsim 3.2$ for $\tan \beta =45$. 
Thus, Figs.~1 and 2 show that the allowed region of $\alpha$ 
 is widened in the presence of $D$-term contributions. 
In order to explicity show this fact, 
 we present the allowed parameter sets of 
 ($\alpha_Y, \alpha_{\rm B-L}$) for fixed $\alpha$ in Fig.~3. 
The point, $\alpha_Y=\alpha_{\rm B-L}=0$, corresponding to 
 the conventional mixed modulus-anomaly mediation 
 is not allowed.

In Tables~\ref{table1} and \ref{table2}, 
 we show some example data 
 of the resultant sparticle mass spectrum and Higgs boson masses 
 for ${\rm sgn}(\mu)>0$. 
Here, the standard model-like Higgs boson mass 
 is evaluated by including one-loop corrections 
 through top and scalar top quarks, 
\bea
 \Delta m_h^2 = \frac{3}{4 \pi^2} y_t^4 v^2 \sin^4 \beta
 \ln \left(\frac{m_{\tilde{t}_1} m_{\tilde{t}_2}}{m_t^2} \right) \;, 
\eea 
which is important to push up the Higgs boson mass 
 so as to satisfy the LEP II experimental bound, $m_h \gtrsim 114$ GeV. 
As can be understood from the RGEs and 
 the soft SUSY breaking parameters at the GUT scale 
 presented in the previous section, 
 the resultant soft SUSY breaking parameters are 
 proportional to $M$.
Thus, as we take $M$ larger with 
 fixed ($\alpha, \alpha_Y, \alpha_{\rm B-L}$), 
 sparticles become heavier and, 
 accordingly, Higgs boson masses become larger. 
In the first column in Table~\ref{table1}, 
 the LSP neutralino is wino-like as the same as 
 in the pure AMSB scenario, 
 while bino-like in the other columns. 
In the last two columns, 
 the LSP neutralino well degenerates with the next LSP sfermion. 
Depending on values of $\alpha_Y$ and $\alpha_{\rm B-L}$, 
 stau or stop can be the next LSP. 
This shows remarkable effects due to the $D$-term contributions. 

The LSP neutralino is a good candidate for the dark matter 
 in cosmology \cite{DM}. 
For small $\tan \beta$, 
 if the LSP neutralino is bino-like, 
 its annihilation processes are dominated by p-wave,  
 and are not so effective that the neutralino relic density 
 tends to over-close the present universe. 
If the LSP neutralino well degenerates with the next LSP sfermions, 
 its co-annihilation process with the next LSP plays an important 
 role to make neutralino annihilation processes effective. 
Our results show that this case is possible 
 due to the effects of the $D$-term contributions 
 on the sfermion masses. 

In the first two columns in Table~\ref{table2}, 
 the lightest neutralino is Higgsino-like, 
 while bino-like in the last two columns. 
For $\alpha \gtrsim 3$, 
 we found that the LSP is stau and this region 
 is cosmologically disfavored. 
Light Higgs boson masses shown in the Table indicate 
 that, in the case of large $\tan \beta$ and small $\alpha$, 
 it is difficult to achieve 
 the correct electroweak symmetry breaking.  

Finally we show sparticle mass spectrum as a function of 
 the parameters of the $D$-terms. 
Fig.~4 (a) shows several sparticle masses as a function of $\alpha_Y$ 
 in the case of $\alpha=5$, $\tan \beta =10$, $M=110$ GeV 
 and $\alpha_{\rm B-L}$ fixed to be $\alpha_{\rm B-L}= \alpha_Y$. 
This figure includes the sparticle masses presented 
 in the second and third columns in Table 1. 
The point of $\alpha_Y=0$ corresponds to 
 the sparticle mass spectrum 
 in the conventional mixed modulus-anomaly mediation. 
We can see that the $D$-term contributions dramatically change 
 the resultant sparticle masses, in particular slepton masses, 
 from those in the conventional mixed modulus-anomaly mediation. 
For $\alpha_Y < -2.5$, lighter stau is mostly left-handed stau, 
 while mostly right-handed stau for $\alpha_Y > -2.5$. 
As discussed above, there exists the parameter region 
 where lighter stau or stop degenerates with the LSP neutralino. 
The case of $\alpha=5$, $\tan \beta =45$, $M=110$ GeV 
 and $\alpha_{\rm B-L}$ fixed as $\alpha_{\rm B-L}= -\alpha_Y$ 
 is depicted in Fig.~4 (b). 
This includes the results in the last two columns in Table 2. 
Sparticle masses moderately depends on $\alpha_Y$ in this case.

\section{Summary and discussion}
We have extended the mixed modulus-anomaly mediation
 so as to include $D$-term contributions 
 from $U(1)_Y$ and $U(1)_{\rm B-L}$. 
Such $D$-term contributions can generically be expected 
 when we consider some grand unified theory  
 based on a higher rank gauge group such as SO(10). 
We have evaluated soft SUSY breaking terms 
 and obtained various sparticle mass spectra 
 for various input values of 
 ($\alpha$, $\alpha_Y$, $\alpha_{\rm B-L}$), 
 that are different from those obtained 
 in the conventional mixed modulus-anomaly mediation. 
Especially, we have found that 
 the LSP neutralino can be various types
 such as wino-like, Higgsino-like and bino-like. 
In addition, stau or stop can be the next LSP 
 with degenerate masses with bino-like neutralino 
 due to the $D$-term contributions. 
This indicates that the co-annihilation channel 
 can be opened up, when we consider the dark matter physics 
 for the bino-like LSP neutralino. 
Evaluating the dark matter relic density in our scenario 
 is an interesting subject. 
We leave this for future works. 

Non-zero $D$-term of $U(1)_{\rm B-L}$ has 
 further phenomenological importance, that is, 
 new flavor violating effects can be generated through it. 
In the presence of the $D$-term of $U(1)_{\rm B-L}$, 
 off-diagonal elements of the slepton mass squared matrix 
 can be generated \cite{Ibe:2004tg}, 
\bea
(\Delta m_{\tilde{\ell}}^2)_{ij} = \frac{1}{8 \pi^2}
(Y_\nu^\dag Y_\nu)_{ij} \;D_{\rm B-L} \;.
\eea
 where $Y_\nu$ is the neutrino Dirac Yukawa coupling matrix. 
Depending on the Yukawa coupling matrix and 
 the value of $D_{\rm B-L}$, 
 the lepton flavor violating (LFV) processes may be sizable. 
The off-diagonal elements can also be induced by RGEs 
 through the neutrino Yukawa coupling as usually discussed \cite{Hisano}. 
Taking these contributions all together, 
 analyzing the LFV processes in our scenario is worth investigating. 
In this analysis, concrete information 
 about neutrino Yukawa coupling is necessary \cite{LFV}.

\section*{Acknowledgments}
We would like to thank Ken-ichi Okumura and Shigeki Matsumoto 
 for useful discussions. 
The work of T.F. and N.O. is supported in part by 
 the Grant-in-Aid for Scientific Research from the Ministry 
 of Education, Science and Culture of Japan (\#16540269, \#15740164).
The work of T.K. is supported by the Research Fellowship 
 of the Japan Society for the Promotion of Science (\#7336).
T.F. and T.K. thank to the Theory Division at KEK for hospitality.


\newpage
\begin{figure}[htbp,width=10cm,height=6cm]
\begin{center}
\subfigure[$(\alpha,~\alpha_Y)$.]
{\includegraphics[height=4.5cm]{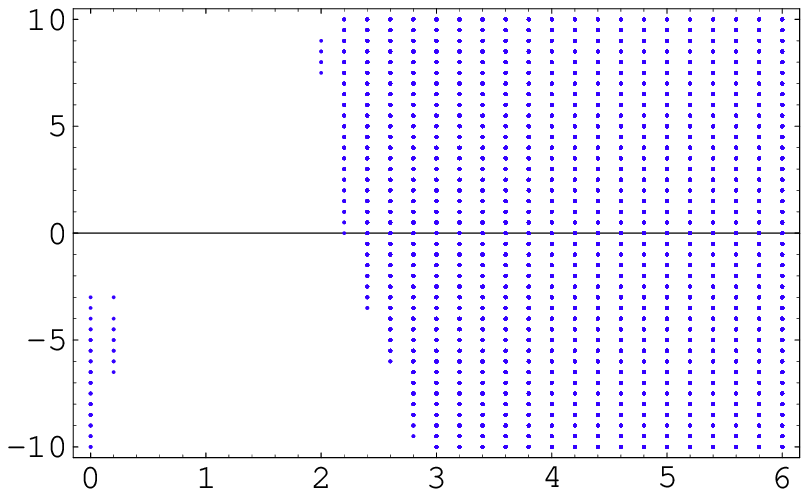}\label{Fig1a}}
\subfigure[$(\alpha,~\alpha_{\rm B-L})$.]
{\includegraphics[height=4.5cm]{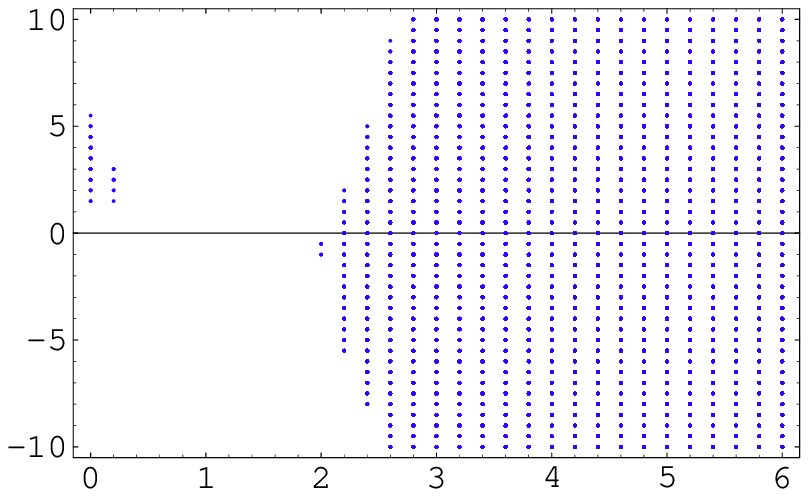}\label{Fig1b}}
\end{center}
\caption{
The allowed parameter set 
 which provides all the sfermion squared masses positive  
 and the correct electroweak symmetry breaking 
 in the case of $\tan \beta=10$ and $M=500$ GeV.}
\vspace{1cm}
\end{figure}
\begin{figure}[htbp,width=10cm, height=6cm]
\begin{center}
\subfigure[$(\alpha,~\alpha_Y)$.]
{\includegraphics[height=4.5cm]{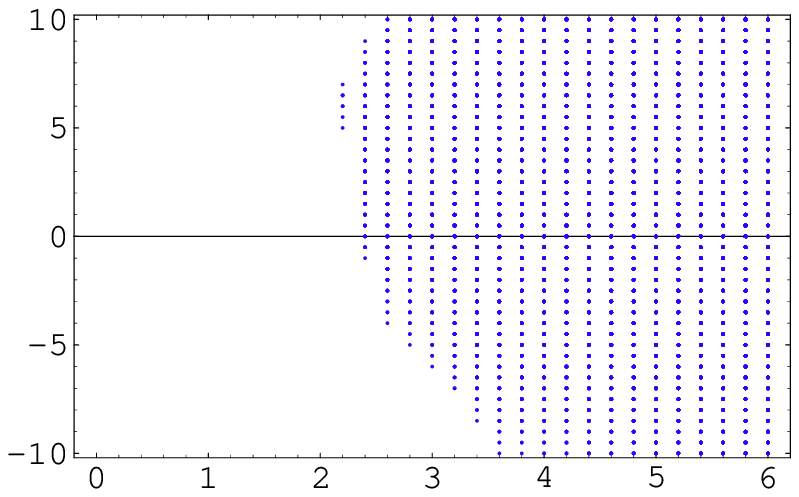}\label{Fig2a}}
\subfigure[$(\alpha,~\alpha_{\rm B-L})$.]
{\includegraphics[height=4.5cm]{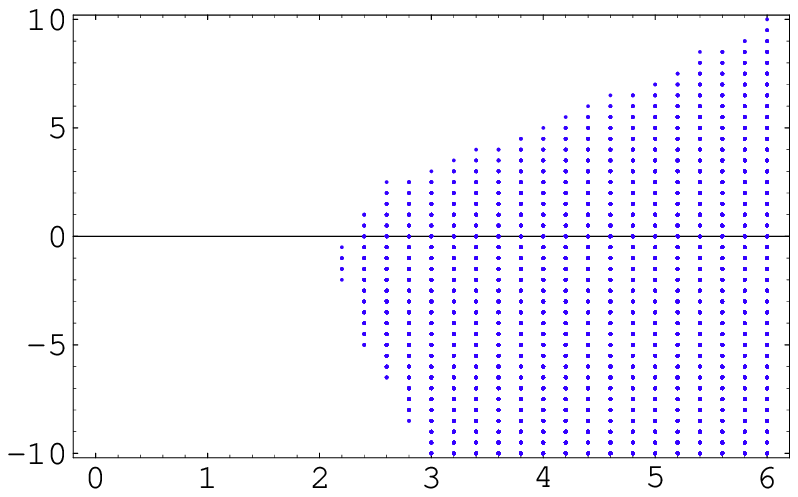}\label{Fig2b}}
\end{center}
\caption{
The allowed parameter set 
 in the case of $\tan \beta=45$ and $M=500$ GeV.}
\end{figure}
%
\newpage
\begin{figure}[ht,width=10cm,height=5cm]
\begin{center}
\subfigure[$(\alpha_Y,~\alpha_{\rm B-L})$.]
{\includegraphics[height=4.5cm]{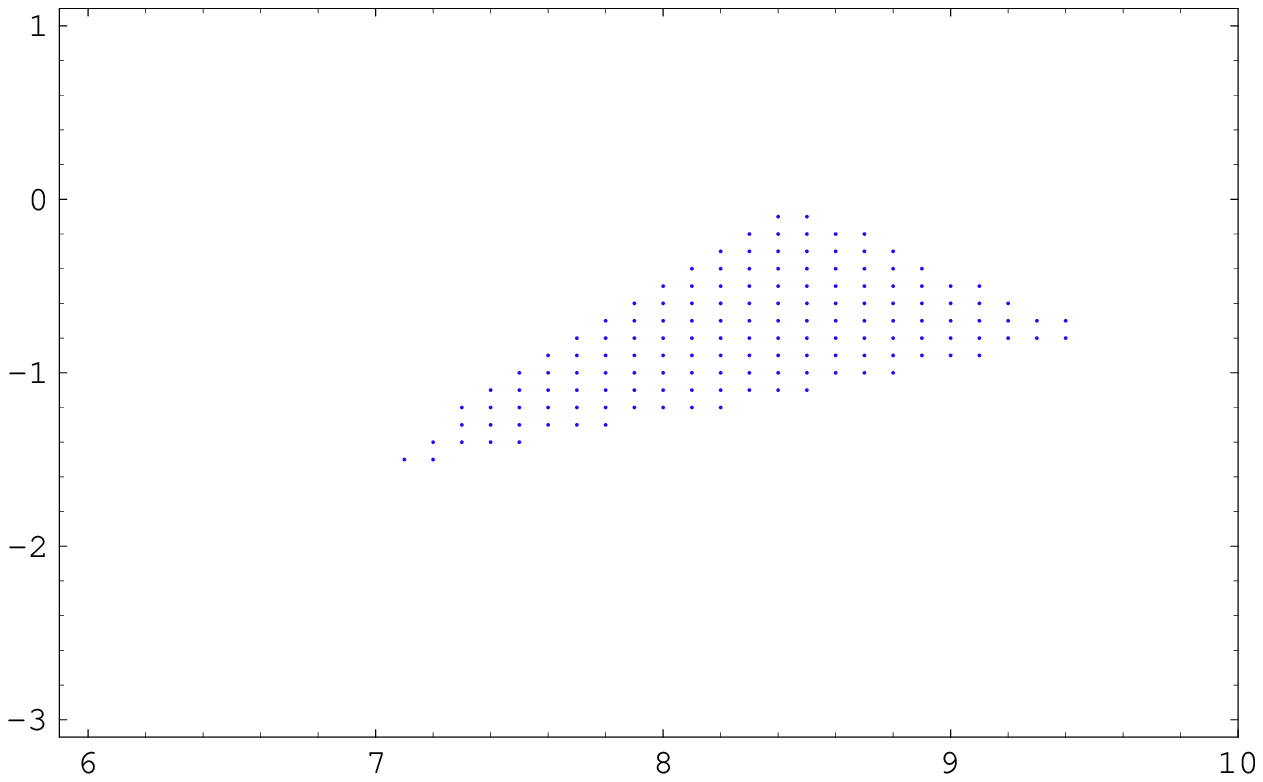}\label{Fig3a}}
\subfigure[$(\alpha_Y,~\alpha_{\rm B-L})$.]
{\includegraphics[height=4.5cm]{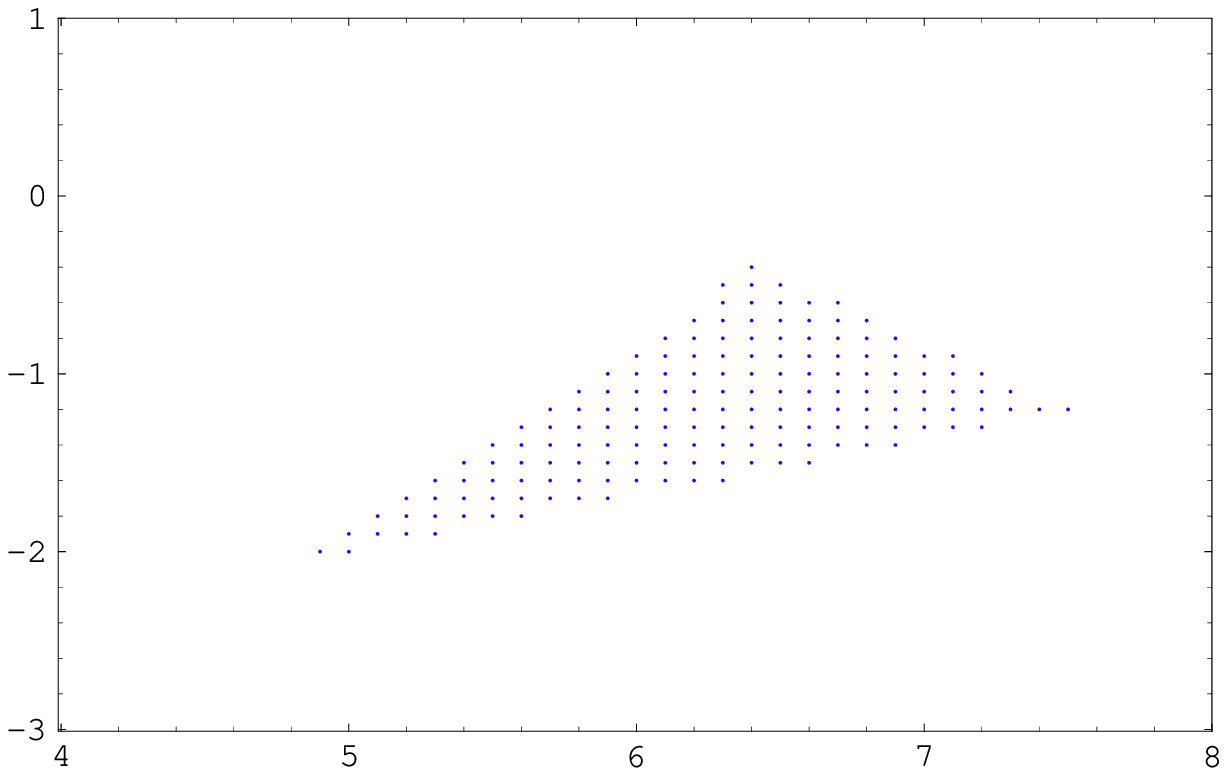}\label{Fig3b}}
\end{center}
\caption{
The allowed parameter set 
 which provides all the sfermion squared masses positive  
 and the correct electroweak symmetry breaking
 in the case of $\alpha=2$, $\tan \beta=10$ and $M=500$ GeV, 
 and (b) $\alpha=2.2$, $\tan \beta=45$ and $M=500$ GeV.
}
\end{figure}
%
\begin{figure}[hb,width=10cm, height=5cm]
\begin{center}
\subfigure[$(\alpha_Y,~{\rm sparticle~mass})$.]
{\includegraphics[height=4.5cm]{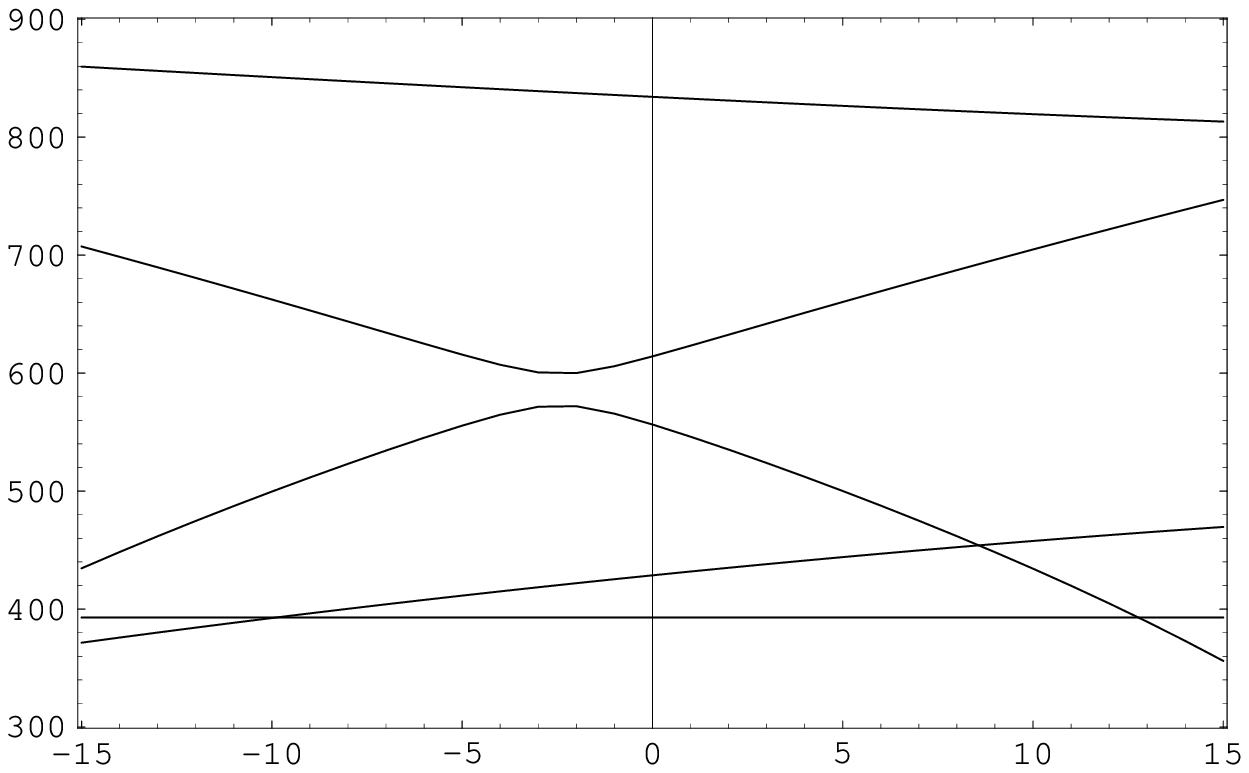}\label{Fig4a}}
\subfigure[$(\alpha_Y,~{\rm sparticle~mass})$.]
{\includegraphics[height=4.5cm]{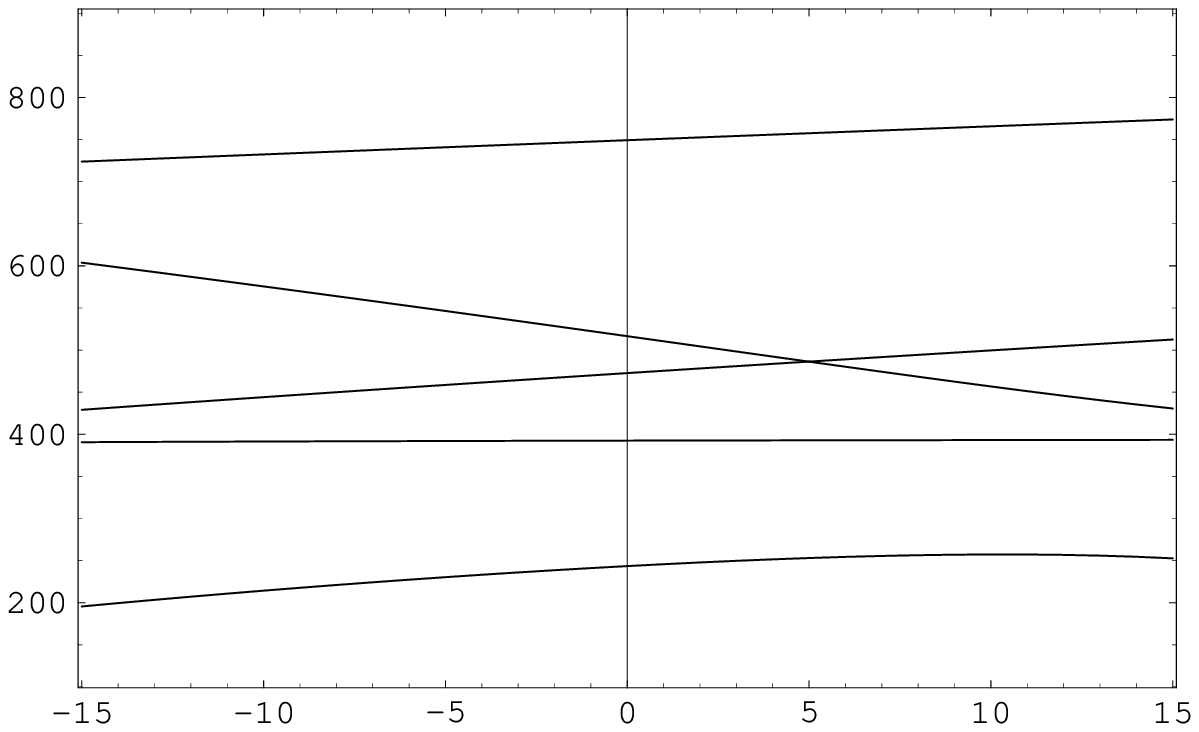}\label{Fig4b}}
\end{center}
\caption{
Sparticle mass spectrum (in units of GeV) as a function of $\alpha_Y$. 
The figure (a) shows the result in the case of 
 $\alpha=5$, $\tan \beta=10$, $M=110$ GeV and 
 $\alpha_{\rm B-L}$ fixed to be $\alpha_{\rm B-L}= \alpha_Y$. 
Each plot corresponds to $m_{\tilde{t}_2}$, 
 $m_{\tilde{\tau}_2}$, 
 $m_{\tilde{\tau}_1}$, 
 $m_{\tilde{t}_1}$ and $m_{\tilde{\chi}_1^0}$, 
 respectively, 
 from top to bottom at $\alpha_Y=0$. 
In the figure (b), $\alpha=5$, $\tan \beta=45$, $M=110$ GeV 
 and $\alpha_{\rm B-L}$ has been taken 
 to be $\alpha_{\rm B-L} = -\alpha_Y$.
Each plot corresponds to 
 $m_{\tilde{t}_2}$, 
 $m_{\tilde{\tau}_2}$, 
 $m_{\tilde{t}_1}$, 
 $m_{\tilde{\chi}_1^0}$ and $m_{\tilde{\tau}_1}$, respectively, 
 from top to bottom at $\alpha_Y=0$. }
\vspace{-2cm}
\end{figure}
\newpage
\begin{center}
\begin{table}[h]
\centering
\begin{tabular}{|c|c|c|c|c|}
\hline
$M$ [GeV]                   & 250
                            & 110
                            & 110
                            & 120  \\
$\alpha $                     & 0
                              & 5
                              & 5
                              & 5  \\
$(\alpha_Y, \alpha_{\rm B-L})$  & ($-6$, $3$)
                                & ($0$, $0$)
                                & ($12.5$, $12.5$) 
                                & ($-12$, $6$)    \\
\hline
$m_{\tilde{\chi}^0_{1,2,3,4}}$     &  99.0, 351, 568, 576
                         & 393,   491, 712, 730
                         & 393,   491, 714, 731
                         & 429,   535, 740, 759 \\
$m_{\tilde{\chi}^{\pm}_{1,2}}$ & 99.0, 575
                               & 490, 729
                               & 490, 730
                               & 534, 758 \\
$m_{\tilde{g}}$          & 902 
                         & 940
                         & 940
                         & 1030 \\
\hline
$m_{{\tilde{e},\tilde{\mu}}_{1,2}}$ 
                          & 178, 219
                          & 575, 618
                          & 419, 732
                          & 646, 697 \\
$m_{\tilde{\tau}_{1,2}}$  & 161, 225
                          & 556, 614
                          & 397, 726
                          & 628, 692 \\
\hline
$m_{{\tilde{u},\tilde{c}}_{1,2}}$  
                         & 784, 826
                         & 911, 943
                         & 915, 940
                         & 971, 1020 \\
$m_{\tilde{t}_{1,2}}$    & 569, 774
                         & 429, 834
                         & 464, 816
                         & 434, 891 \\
\hline
$m_{{\tilde{d},\tilde{s}}_{1,2}}$  
                         & 829, 930
                         & 907, 946
                         & 918, 934
                         & 1020, 1030\\
$m_{\tilde{b}_{1,2}}$    & 733, 918
                         & 763, 891
                         & 729, 912
                         & 826, 1000\\
\hline
$m_h$                    & 116
                         & 114
                         & 115
                         & 115 \\
$m_H$                    & 373
                         & 911
                         & 913
                         & 936 \\
$m_A$                    & 373
                         & 911
                         & 913
                         & 936 \\
$m_{H^{\pm}}$            & 382
                         & 914
                         & 917
                         & 939 \\
\hline
\end{tabular}
\caption{
Sparticle and Higgs boson mass spectra (in units of GeV) 
 in the case of $\tan \beta=10$.}
\label{table1}
\end{table}
\end{center}
\begin{center}
\begin{table}[h]
\centering
\begin{tabular}{|c|c|c|c|c|}
\hline
$M$ [GeV]                   & 450
                            & 280
                            & 110
                            & 110  \\
$\alpha $                     & 2.5
                              & 3
                              & 5
                              & 5  \\
$(\alpha_Y, \alpha_{\rm B-L})$  & ($-3$, $-4.5$)
                                & ($-3$, $-4.5$)
                                & (0,0) 
                                & ($6$,$-6$)    \\
\hline
$m_{\tilde{\chi}^0_{1,2,3,4}}$           & 174, 181, 1130, 1135
                               & 480, 499, 714, 775
                               & 392, 486, 633, 656
                               & 393, 489, 657, 677 \\
$m_{\tilde{\chi}^{\pm}_{1,2}}$ & 178, 1130
                               & 486, 774
                               & 485, 655
                               & 488, 676 \\
$m_{\tilde{g}}$          &  1110
                         &  958
                         &  940
                         &  940 \\
\hline
$m_{{\tilde{e},\tilde{\mu}}_{1,2}}$ 
                          & 637, 1440
                          & 614, 947 
                          & 575, 618
                          & 583, 585  \\
$m_{\tilde{\tau}_{1,2}}$  & 263, 118
                          & 420, 716 
                          & 243, 516
                          & 254, 480  \\
\hline
$m_{{\tilde{u},\tilde{c}}_{1,2}}$  
                         & 1000, 1230
                         & 902, 1010
                         & 911, 943
                         & 920, 950  \\
$m_{\tilde{t}_{1,2}}$    & 343, 948
                         & 387, 787
                         & 473, 749
                         & 489, 759  \\
\hline
$m_{{\tilde{d},\tilde{s}}_{1,2}}$  
                         & 948, 1230
                         & 879, 1020
                         & 907, 946
                         & 882, 953  \\
$m_{\tilde{b}_{1,2}}$    & 504, 888
                         & 531, 723
                         & 570, 709 
                         & 546, 706  \\
\hline
$m_h$                    & 115
                         & 114
                         & 115
                         & 116 \\
$m_H$                    & 203
                         & 154
                         & 112
                         & 272 \\
$m_A$                    & 203
                         & 154
                         & 112
                         & 272  \\
$m_{H^{\pm}}$            & 219
                         & 173
                         & 138
                         & 284 \\
\hline
\end{tabular}
\caption{
Sparticle and Higgs boson mass spectra (in units of GeV) 
 in the case of $\tan \beta=45$.}
\label{table2}
\vspace{-2cm}
\end{table}
\end{center}


\begin{thebibliography}{999}

\bibitem{SUSY}
For a general review of supersymmetry, 
 see, for example, 
 H.~P.~Nilles, Phys.\ Rept.\  {\bf 110} (1984) 1, 
 and references therein.

\bibitem{Luty:2005sn}
For a review of supersymmetry breaking, 
 see, for example, M.~A.~Luty,
 arXiv:hep-th/0509029, and references therein.

\bibitem{AMSB1}
  L.~Randall and R.~Sundrum,
  Nucl.\ Phys.\ B {\bf 557}, 79 (1999)
  [arXiv:hep-th/9810155];

\bibitem{AMSB2}
  G.~F.~Giudice, M.~A.~Luty, H.~Murayama and R.~Rattazzi,
  JHEP {\bf 9812}, 027 (1998)
  [arXiv:hep-ph/9810442].

\bibitem{tree1}
 I.~Jack and D.~R.~T.~Jones,
 Phys.\ Lett.\ B {\bf 482}, 167 (2000)
 [arXiv:hep-ph/0003081];
%
 N.~Arkani-Hamed, D.~E.~Kaplan, H.~Murayama and Y.~Nomura,
 JHEP {\bf 0102}, 041 (2001)
 [arXiv:hep-ph/0012103].

\bibitem{tree2}
  N.~Kitazawa, N.~Maru and N.~Okada,
  Phys.\ Rev.\ D {\bf 62}, 077701 (2000)
  [arXiv:hep-ph/9911251]; 
  Nucl.\ Phys.\ B {\bf 586}, 261 (2000)
  [arXiv:hep-ph/0003240]; 
  Phys.\ Rev.\ D {\bf 63}, 015005 (2001)
  [arXiv:hep-ph/0007253].

\bibitem{PR}
A.~Pomarol and R.~Rattazzi,
JHEP {\bf 9905}, 013 (1999) 
 [arXiv:hep-ph/9903448].

\bibitem{quantum}
Z.~Chacko, M.~A.~Luty, I.~Maksymyk and E.~Ponton,
 JHEP {\bf 0004}, 001 (2000) 
 [arXiv:hep-ph/9905390];
E.~Katz, Y.~Shadmi and Y.~Shirman,
 JHEP {\bf 9908}, 015 (1999) 
 [arXiv:hep-ph/9906296]; 
B.~C.~Allanach and A.~Dedes,
 JHEP {\bf 0006}, 017 (2000) 
 [arXiv:hep-ph/0003222];
%
D.~E.~Kaplan and G.~D.~Kribs,
 JHEP {\bf 0009}, 048 (2000)
 [arXiv:hep-ph/0009195];
%
Z.~Chacko and M.~A.~Luty, 
JHEP {\bf 0205}, 047 (2002)
 [arXiv:hep-ph/0112172]; 
%
Z.~Chacko and E.~Ponton,
Phys.\ Rev.\ D {\bf 66}, 095004 (2002) 
 [arXiv:hep-ph/0112190]; 
%
N.~Okada, 
Phys.\ Rev.\ D {\bf 65}, 115009 (2002)
 [arXiv:hep-ph/0202219]; 
%
A.~E.~Nelson and N.~T.~Weiner, 
 arXiv:hep-ph/0210288; 
%
O.~C.~Anoka, K.~S.~Babu and I.~Gogoladze,
Nucl.\ Phys.\ B {\bf 686}, 135 (2004) 
 [arXiv:hep-ph/0312176].


\bibitem{Kitano:2004zd}
  R.~Kitano, G.~D.~Kribs and H.~Murayama,
  Phys.\ Rev.\ D {\bf 70}, 035001 (2004)
  [arXiv:hep-ph/0402215].

\bibitem{Kachru:2003aw}
  S.~Kachru, R.~Kallosh, A.~Linde and S.~P.~Trivedi,
  Phys.\ Rev.\ D {\bf 68}, 046005 (2003)
  [arXiv:hep-th/0301240].

\bibitem{Choi:2004sx}
  K.~Choi, A.~Falkowski, H.~P.~Nilles, M.~Olechowski and S.~Pokorski,
  JHEP {\bf 0411}, 076 (2004)
  [arXiv:hep-th/0411066];
  K.~Choi, A.~Falkowski, H.~P.~Nilles and M.~Olechowski,
  Nucl.\ Phys.\ B {\bf 718}, 113 (2005)
  [arXiv:hep-th/0503216];
  K.~Choi, K.~S.~Jeong and K.~i.~Okumura,
  JHEP {\bf 0509}, 039 (2005)
  [arXiv:hep-ph/0504037];
  A.~Falkowski, O.~Lebedev and Y.~Mambrini,
  JHEP {\bf 0511}, 034 (2005)
  [arXiv:hep-ph/0507110];

\bibitem{KKLT2}
  M.~Endo, M.~Yamaguchi and K.~Yoshioka,
  Phys.\ Rev.\ D {\bf 72}, 015004 (2005)
  [arXiv:hep-ph/0504036].


\bibitem{KKLT3}
K.~Choi, K.~S.~Jeong, T.~Kobayashi and K.~i.~Okumura,
  Phys.\ Lett.\ B {\bf 633}, 355 (2006)
  [arXiv:hep-ph/0508029]; 
R.~Kitano and Y.~Nomura,
  Phys.\ Lett.\ B {\bf 631}, 58 (2005)
  [arXiv:hep-ph/0509039]; 
  Phys.\ Rev.\ D {\bf 73}, 095004 (2006)
  [arXiv:hep-ph/0602096]. 


\bibitem{Kaku:1978nz}
  M.~Kaku, P.~K.~Townsend and P.~van Nieuwenhuizen,
  Phys.\ Rev.\ D {\bf 17} (1978) 3179;
  W.~Siegel and S.~J.~J.~Gates,
  Nucl.\ Phys.\ B {\bf 147}, 77 (1979);
  E.~Cremmer, S.~Ferrara, L.~Girardello and A.~Van Proeyen,
  Nucl.\ Phys.\ B {\bf 212}, 413 (1983);
  S.~Ferrara, L.~Girardello, T.~Kugo and A.~Van Proeyen,
  Nucl.\ Phys.\ B {\bf 223}, 191 (1983); 
  T.~Kugo and S.~Uehara,
  Nucl.\ Phys.\ B {\bf 222}, 125 (1983); 
  Nucl.\ Phys.\ B {\bf 226}, 49 (1983);
  Prog.\ Theor.\ Phys.\  {\bf 73}, 235 (1985).

\bibitem{method}
  G.~F.~Giudice and R.~Rattazzi,
  Nucl.\ Phys.\ B {\bf 511}, 25 (1998)
  [arXiv:hep-ph/9706540]; 
%
  N.~Arkani-Hamed, G.~F.~Giudice, M.~A.~Luty and R.~Rattazzi,
  Phys.\ Rev.\ D {\bf 58}, 115005 (1998)
  [arXiv:hep-ph/9803290].

\bibitem{RGE}
 D.~J.~Castano, E.~J.~Piard and P.~Ramond,
  Phys.\ Rev.\ D {\bf 49}, 4882 (1994)
  [arXiv:hep-ph/9308335].

\bibitem{DM}
 For a review, see, for example, 
 G.~Jungman, M.~Kamionkowski and K.~Griest,
  Phys.\ Rept.\  {\bf 267}, 195 (1996)
  [arXiv:hep-ph/9506380], 
 and references therein. 


\bibitem{Ibe:2004tg}
  M.~Ibe, R.~Kitano, H.~Murayama and T.~Yanagida,
  Phys.\ Rev.\ D {\bf 70}, 075012 (2004)
  [arXiv:hep-ph/0403198].

\bibitem{Hisano} 
For a recent review, see, for example, 
J.~Hisano, 
Nucl.\ Phys.\ Proc.\ Suppl.\  {\bf 111}, 178 (2002)
[arXiv:hep-ph/0204100], and references therein. 

\bibitem{LFV} 
 For an analysis of LFV processes 
 by using concrete Yukawa coupling matrices 
 predicted in the minimal SUSY SO(10) GUT 
 with minimal supergravity mediation, 
 see, for example, 
  T.~Fukuyama, T.~Kikuchi and N.~Okada,
  Phys.\ Rev.\ D {\bf 68}, 033012 (2003)
  [arXiv:hep-ph/0304190].


\end{thebibliography}
\end{document}